\documentclass[twocolumn,showpacs,preprintnumbers,amsmath,amssymb]{revtex4}

\usepackage{graphicx}
\usepackage{dcolumn}
\usepackage{bm}

\begin{document}

\title{Combined scanning force microscopy and scanning tunneling spectroscopy of an electronic nano-circuit at very low temperature}

\author{J. Senzier, P. S. Luo and H. Courtois$^*$} 
\affiliation{Centre de Recherches sur les Tr\`es Basses Temp\'eratures - C.N.R.S.  and Universit\'e Joseph Fourier, 25 Avenue des Martyrs, 38042 Grenoble, France.\\
*also at Institut Universitaire de France}

\date{\today}

\begin{abstract}
We demonstrate the combination of scanning force microscopy and scanning tunneling spectroscopy in a local probe microscope operating at very low temperature (60 mK). This local probe uses a quartz tuning fork ensuring high tunnel junction stability. We performed the spatially-resolved spectroscopic study of a superconducting nano-circuit patterned on an insulating substrate. Significant deviations from the theoretical prediction are observed.
\end{abstract}

                              
\maketitle

Scanning Tunneling Microscopy (STM) is a powerful tool for imaging surfaces down to the atomic scale. It also allows to measure the electronic Local Density Of States (LDOS) with an accuracy of $2k_B T$ given by the thermal smearing. This so-called Scanning Tunneling Spectroscopy (STS) technique has been used in many fields of condensed matter physics. Cryogenic environment is a key to reach the best spectroscopic energy resolution and access the fundamental electronic properties of quantum mechanics-driven nanostructures. Nevertheless, the application of STS in the field of mesoscopic devices and nano-circuits is still in its infancy. The main reason is that a STM needs a fully metallic surface to be operated safely, which is not the case of nano-circuits patterned on insulating substrates.

On the other hand, the Atomic Force Microscopy (AFM) technique \cite{Giessbl} is capable of imaging any kind of surface with an excellent resolution, although no information can be gained on the electronic LDOS. A combination of both modes of microscopy in a versatile instrument, namely a cryogenic AFM-STM with the STS capability, is therefore of particular interest. As the same tip is used for both microscopy, a metallic nano-circuit can be located with the force microscopy before being probed in the tunneling mode. The local spectroscopy of the nano-structure is then possible like in an usual STS experiment. 

The STS is a very demanding technique in terms of junction stability. The use of a metallized cantilever as the force sensor is therefore prohibited, since its natural flexibility would prevent a sufficient tunnel junction stability. An alternative consists in the use of a quartz tuning fork (used in watches), which was proposed in the framework of Scanning Near-field Acoustic Microscopy \cite{Gunther} and since then applied in many kinds of local probes \cite{Karrai,Grober,RSI-Klaus,APL-Seo}. A tuning fork-based AFM probe brings the advantage of its natural stiffness at low frequency, which is a prerequisite for the STS. Moreover, the oscillation of a tuning fork can be detected with electrical means, avoiding any optical setup. Finally, a tuning fork dissipates a very small amount of power. This makes the tuning fork-based AFM the choice candidate for AFM-STM experiments at very low temperature.

In this letter, we report on the characteristics and performances of a combined AFM-STM operating at very low temperature down to 60 mK in a dilution refrigerator. We demonstrate its application to the study of the local electronic properties of a superconducting submicron wire patterned on a sapphire substrate. Significant deviations from the BCS prediction are observed, which are related to the very small thickness of the wire.

Our AFM-STM is an evolution of a home-made very low temperature STM \cite{RSI-STM,Manips-STM}. The microscope is thermalized on the top plate of an upside-down dilution refrigerator with a base temperature of 60 mK. Every electrical connection entering the microscope's hermetic calorimeter is now filtered with radio-frequency filters made of 20 cm-long high capacitance micro-coaxial wires \cite{Kumar}. 

We are using commercial quartz tuning forks resonating at 32768 Hz. To mount a probe, we first cut the tuning fork packaging just above the fork prongs end. Keeping the tuning fork box avoids possible damage to the tuning fork itself and provides a good electrical shield for the electrical measurements. Then we glue an electrochemically-etched tungsten tip with a small amount of conductive epoxy at the end of the top prong, see Fig. 1 inset. The metallic contact between the tip and one tuning fork electrode enables us to use the latter as the tunnel current collector while the other electrode is used for the AFM signal measurement. We chose to mount the tip perpendicular to the tuning fork prongs, so that tip-sample interaction forces are probed, not shear forces. 

\begin{figure}[ht]
\centering
\includegraphics{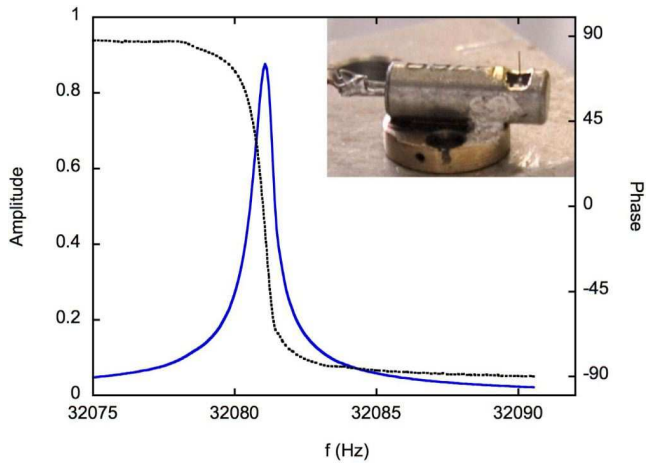}
\caption{\label{tuningfork} (Color online) Plot of the phase (dotted line) and amplitude (full line) of the AFM-STM probe response close to its resonance. The measurement is performed at 4 K with a mechanical excitation through the scanner tube. Inset: Photograph of the probe made of a tungsten tip glued with silver epoxy on one electrode of the quartz tuning fork. The cylindrical tuning fork packaging (length 8 mm) was cut just above the fork prongs prior to gluing.}
\end{figure}

After fabrication, the tuning fork probe is fixed on the microscope scanner tube. It is excited mechanically by adding a small a.c. component to the high voltage applied to the tube's inner electrode. Although the tube response is attenuated by about two decades in amplitude in the relevant frequency range (30-32 kHz), this method avoids the use of an additional piezo-electric element and enables us to measure simultanously the tunnel current and the AFM signal. Compared to a bare tuning fork, the resonance frequency shift and the quality factor of a tuning-fork probe depend strongly on the amount of glue and on the geometry of the tip gluing. Figure 1 displays a typical frequency response of an AFM-STM probe at cryogenic temperature. The resonance frequency is shifted down by about $700 Hz$ while the quality factor of about 50000 remains quite good. 

As the tip approaches the sample surface, the tuning fork oscillation is influenced by the tip-sample interaction forces. In the small amplitude limit, the resonance frequency is shifted proportionally to the force gradient \cite{Giessbl,Rychen-ASS}. Figure 2a shows a typical dependence of the frequency shift $\Delta f$ as a function of the tip-sample distance. The zero distance reference has been defined with an estimated accuracy of 0.3 nm as the position where a tunnel resistance of $1 G\Omega$ is reached (at a 0.5 Volt bias). The oscillation amplitude of 0.8 nm is close to the small amplitude limit, so that the resonance frequency shows a clear negative shift regime and exceeds - 1 Hz at a distance of about 3 nm. For the sake of operation reliability, we chose to operate in the positive shift region which features a steep evolution of the frequency shift as a function of distance of the order of 2 Hz per nanometer. 

\begin{figure}[ht]
\includegraphics{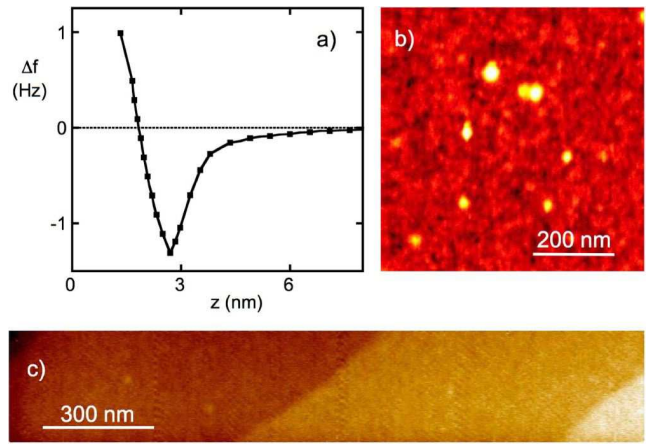}
\caption{\label{AFM} (Color online) (a) Frequency shift $\Delta f$ as a function of the tip-sample distance on a Nb-Au film at 150 mK. (b) $0.7 \times 0.7 \mu m^2$ very low temperature AFM image of a silica surface with adsorbed Au nano-particles of 5 nm diameter: $\Delta f = 0.4 Hz$ and T = 100 mK, z-scale is 6 nm. (c) $1600 \times 300 \ n m^2$ AFM image of atomic steps on an annealed R-plane sapphire wafer: $\Delta f = 0.2 Hz$ and T = 90 mK, z-scale is 3 nm.}
\end{figure}

As for AFM imaging, we use a Phase Locked Loop electronics that maintains the excitation frequency precisely at the resonance, while an adjustable gain amplifier also maintains the oscillation amplitude constant. In this Frequency Modulation mode (FM-AFM), the measured frequency shift feeds the distance regulation electronics that operate with a setpoint of 0.15 to 0.5 Hz. As for test images, Fig. 2b shows an image of a reference sample made of a silanized silica surface with adsorbed colloidal Au nano-particles of 5 nm diameter. Dots appear with a height of 5 nm and a full width at half maximum of about 35 nm. This apparent size increase shows that our lateral resolution is limited by a tip radius of about 15 nm. In the vertical direction, we achieve routinely atomic resolution. As an example, the Fig. 2c image of a sapphire R-plane substrate annealed for 1h in air at 1100$^\circ$C displays regular atomic steps (step height of 0.34 nm). The measured corrugation on a given atomic plane is 0.1 nm, which demonstrates the excellent vertical resolution of the set-up. 

We have achieved very low temperature AFM-STS experiments on submicron Nb wires patterned on such a sapphire substrate. An epitaxial Nb layer of 15 nm was deposited in UHV at 500$^\circ$C and covered by a thin Si (5 nm) capping layer \cite{APL-Bouchiat}. We used e-beam lithography and Reactive Ion Etching (RIE) to pattern a Nb meander line. Our process provides a clean Nb surface, although a few small local defects may remain.  Figure 3a shows a 3D representation from an AFM image of a sample at T = 100 mK. A Nb wire with a width of about 300 nm and a height of 8 nm is imaged, in agreement with expectations. The observed r.m.s. corrugation on the wire's surface is about 0.8 nm. After taking several images, we moved the probe over the wire and switched the microscope to the STM mode. At this moment, the tuning fork is no longer excited and the input of the distance regulation is fed with the tunnel current signal. 

\begin{figure}[ht]
\includegraphics{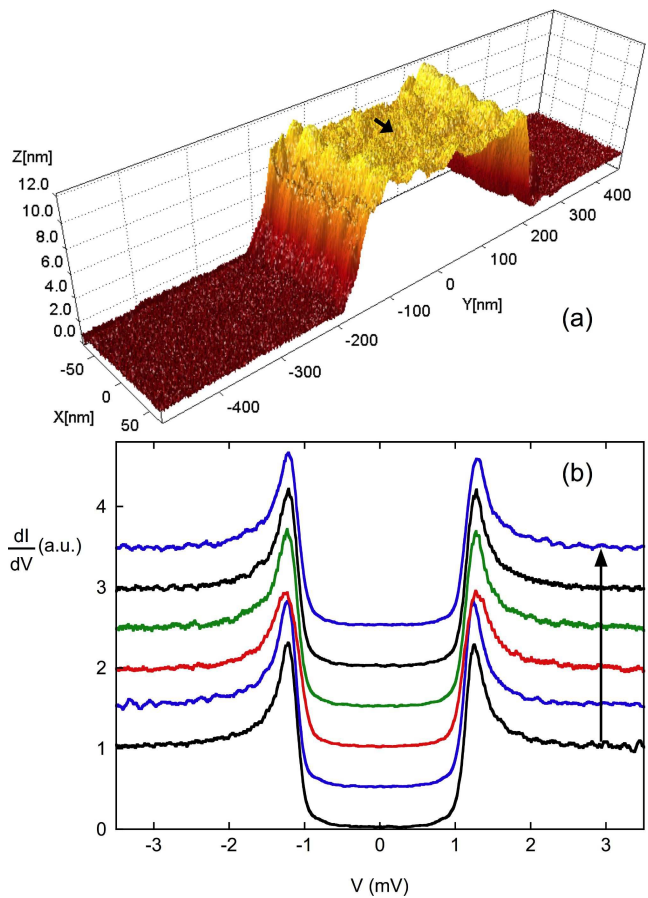}
\caption{\label{Spectroscopie} (Color online) (a) $ 140\times 950 \ nm^2$ AFM image of a 300 nm-wide Nb wire patterned on a sapphire substrate at 100 mK. Frequency shift is $0.15 Hz$, tip oscillation amplitude is 1 nm, scanning speed is 175 nm/s. (b) Differential conductance spectra acquired at 100 mK in units of the high bias conductance equal to 50 M$\Omega$. The spectroscopies from bottom to top were acquired along a $25 nm$-long displacement of the tip on the Nb wire. The arrow superposed on the AFM image gives the position, length and direction of the spectroscopic scan.}
\end{figure}

We performed local spectroscopy measurements by acquiring direct I-V characteristics that were subsequently numerically differentiated in order to access the LDOS. A series of spectroscopies was acquired along the arrow displayed on Fig. 3b and with a 5 nm displacement between every spectroscopic measurement. In this case, displacement and spectroscopic measurements were made with a constant tip-sample distance corresponding to a $50$ $M\Omega$ tunnel barrier. The tunnel resistance could be decreased down to $10$ $M\Omega$ while still keeping a good junction stability. The noise level of the I-V characteristics is due to both the vibration of the tip-sample distance and the measurement noise. The exponential dependence of the tunnel current $I$ with the distance $z$ gives the formula: 
\begin{equation}
\frac{dI}{I}=-\frac{\sqrt{2 m W}}{\hbar}dz,
\end{equation}
$m$ being the electron mass and $W$ the metal work function of about 4 eV. Neglecting the effect of the measurement noise, the measured current noise level of below 0.5$\%$ gives then an upper limit for the peak-peak tip vibration of 0.5 pm. This demonstrates the high stability of the tuning fork-based AFM-STM probe.

The Figure 3b spectra display a clear superconducting gap with a width of $2 \Delta$=$2.32$ meV which is slightly below the gap of bulk Nb (3.0 meV) but consistent with the measured critical temperature of 7.4 K. The expected superconducting coherence length is 9.8 nm. The third spectrum shows slightly smaller coherence peaks than the others. This behavior is presumably related to a local defect that weakens locally the superconductivity. The other spectra are very close to each other and show a smearing that cannot be accounted for by the measured sample temperature (100 mK).

As an energy resolution test, we performed AFM-STM spectroscopy experiments on thick Nb films. We obtained spectra that could be fitted by the Bardeen-Cooper-Schrieffer (BCS) theory and an effective temperature close to 200 mK, in agreement with our recent STS experiments \cite{Manips-STM}. We therefore conclude that our measurements are not affected by thermal smearing or noise and give an accurate measure of the LDOS of the Nb surface. The LDOS thus features smeared peaks and a significant sub-gap density, in clear deviation from the BCS theory. This behavior appears to be linked to the small thickness of the film as the same behavior was observed on unprocessed films of similar thickness. An inverse proximity effect due to the presence of a metallic and non-superconducting Nb oxide layer at the sample surface or at the sapphire-Nb interface is a possible explanation for the observed behaviour which calls for further investigation.

In conclusion, we have developed a very low temperature local probe combining Force Microscopy and Scanning Tunneling Spectroscopy. The high rigidity of the tuning fork used as a force sensor enables us to obtain highly stable measurement conditions as demonstrated by a first spectroscopy experiment on a submicron superconducting wire. This combined AFM-STM technique opens wide perspectives in the study of quantum nano-circuits. This set-up can be used for Scanning Tunneling Potentiometry of a current-biased device. By using a superconductive tip on a metallic sample, out-of-equilibrium effects can be investigated through measurements of the local energy distribution function \cite{PRL-Pothier}.

We thank the CRTBT technical staff, and especially A. GÕerardin, for their various contributions to the design and construction of the microscope set-up. We acknowledge preliminary work by A. Gupta and L. Cr\' etinon and useful discussions with V. Bouchiat, C. Delacour, N. Dempsey, B. Gr\' evin, K. Hasselbach and P. Joyez. Samples were prepared at the Nanofab facility at CNRS Grenoble. This work was funded by the EU NMP2-CT-2003-505587 'SFINx' STREP and by ACN 'AFM-STM-NME' project.

\end{document}